\documentclass[fleqn,usenatbib]{mnras}

\usepackage{newtxtext,newtxmath}

\usepackage[T1]{fontenc}
\usepackage{ae,aecompl}

\usepackage{graphicx}	
\usepackage{amsmath}	
\usepackage{physics}
\usepackage{tabularx}

\newcommand\T{\rule{0pt}{2.6ex}}       
\newcommand\B{\rule[-1.2ex]{0pt}{0pt}} 

\title[Field linkage and magnetic helicity density]{Field linkage and magnetic helicity density}

\author[K. Lund et al.]{K. Lund$^{1}$,
M. Jardine$^{1}$, A. J. B. Russell$^{2}$, J.-F. Donati$^{3}$, R. Fares$^{4}$, C. P. Folsom$^{3}$,
\newauthor S. V. Jeffers$^{5}$, S. C. Marsden$^{6}$, J. Morin$^{7}$, P. Petit$^{4}$ and V. See$^{8}$\\
$^{1}$SUPA, School of Physics and Astronomy, University of St Andrews, North Haugh, St Andrews, KY16 9SS, UK\\
$^{2}$School of Science \& Engineering, University of Dundee, Nethergate, Dundee DD1 4HN, UK\\
$^{3}$IRAP, Universit\'{e} de Toulouse, CNRS, UPS, CNES, 14 Avenue Edouard Belin, 31400, Toulouse, France\\
$^{4}$Physics Department, United Arab Emirates University, P.O. Box 15551, Al-Ain, United Arab Emirates\\
$^{5}$Institut f\"ur Astrophysik, Universit\"at G\"ottingen, Friedrich-Hund-Platz 1, D-37077 G\"ottingen, Germany\\
$^{6}$University of Southern Queensland, Centre for Astrophysics, Toowoomba, QLD, 4350, Australia\\
$^{7}$LUPM, Universit\'e de Montpellier, CNRS, Place Eug\`ene Bataillon, F-34095 Montpellier, France\\
$^{8}$University of Exeter, Department of Physics \& Astronomy, Stocker Road, Devon, Exeter, EX4 4QL, UK\\
}

\date{Accepted XXX. Received YYY; in original form ZZZ}

\pubyear{2020}

\begin{document}
\label{firstpage}
\pagerange{\pageref{firstpage}--\pageref{lastpage}}
\maketitle
\begin{abstract}
The helicity of a magnetic field is a fundamental property that is conserved in ideal MHD. It can be explored in the stellar context by mapping large-scale magnetic fields across stellar surfaces using Zeeman-Doppler imaging. A recent study of 51 stars in the mass range 0.1-1.34 M$_\odot$ showed that the photospheric magnetic helicity density follows a single power law when plotted against the toroidal field energy, but splits into two branches when plotted against the poloidal field energy. These two branches divide stars above and below $\sim$ 0.5 M$_\odot$. We present here a novel method of visualising the helicity density in terms of the linkage of the toroidal and poloidal fields that are mapped across the stellar surface. This approach allows us to classify the field linkages that provide the helicity density for stars of different masses and rotation rates. We find that stars on the lower-mass branch tend to have toroidal fields that are non-axisymmetric and so link through regions of positive and negative poloidal field. A lower-mass star may have the same helicity density as a higher-mass star, despite having a stronger poloidal field. Lower-mass stars are therefore less efficient at generating large-scale helicity.


\end{abstract}




\section{Introduction}

Magnetic helicity is a fundamental property of magnetic fields that measures the amount of linkage and twist of field lines within a given volume. Since it is exactly conserved in ideal MHD and highly conserved for high magnetic Reynolds numbers in general \citep{Woltjer1958,Taylor1974}, helicity is an important factor when attempting to understand how magnetic fields are generated and evolve \citep[e.g.][]{Brandenburg2005,Chatterjee2011,Pipin2019}.  Until recently, this could only be measured for the Sun \citep[e.g. reviews by][]{Demoulin2007,Demoulin2009}. We can, however, now map all three components of the large-scale magnetic field at the surfaces of stars using the spectropolarimetric technique of Zeeman-Doppler imaging \citep{Semel1989}. 

These magnetic field maps now exist for a large enough sample of stars that trends with stellar mass and rotation period have become apparent \citep{donati2009}. In particular, it appears that magnetic fields show different strengths and topologies in the mass ranges above and below $\sim$ 0.5 M$_\odot$, which is believed to correspond to the onset of the transition from partially to fully convective interiors. Rapidly-rotating stars in the mass range above $\sim$ 0.5 M$_\odot$ tend to have fields that are predominantly toroidal \citep{Donati2008b}. The stronger the toroidal field, the more likely it is to be axisymmetric \citep{See2015}. In the mass range below $\sim$ 0.5 M$_\odot$, stars show predominantly axisymmetric poloidal fields. For the lowest masses, however, a bimodal behaviour is found, such that stars may have strong, predominantly axisymmetric poloidal fields, or much weaker, non-axisymmetric poloidal fields \citep{Donati2008b,Morin2008b,donati2009,Morin2010}.

This difference in magnetic fields in stars that are partially or fully convective is also apparent in their photospheric helicity densities. Using observations of 51 stars, \citet{Lund2020} found that the helicity density scales with the toroidal energy according to $|\langle{h\,}\rangle|$ $\propto$ $\langle{\rm{B_{tor}}^2_{}\,\rangle}^{0.86\,\pm\,0.04}$. The scaling with the poloidal energy is more complex, however, revealing two groups with different behaviours. Specifically, stars less massive than $\sim$ 0.5 M$_\odot$ appear to have an excess of poloidal energy when compared to more massive stars with similar helicity densities. It appears that stars with different internal structures and different total magnetic energies may nonetheless generate magnetic fields with the same helicity density at their surfaces. The aim of this paper is to explore the nature of this division and the types of flux linkage that support the measured helicity densities. In order to do that, we have developed a novel method of visualising the linkages of different field components across the surfaces of stars.

\section{Methods}
\label{sec:methods}

\subsection{Poloidal and toroidal magnetic field components}

For the purposes of this paper the stellar magnetic fields discussed will be decomposed into their poloidal and toroidal components: $\boldsymbol{B}=\boldsymbol{B}_{\rm{pol}}+\boldsymbol{B}_{\rm{tor}}$. The poloidal and toroidal fields can be expressed in a general form as \citep[see Appendix III of ][]{Chandrasekhar1961}:
\begin{equation}
\boldsymbol{B}_{\rm{pol}}=\curl{[\curl{[\Phi\hat{r}]}]}, 
\label{eq:Bporiginal}
\end{equation}
\begin{equation}
\boldsymbol{B}_{\rm{tor}}=\curl{[{\Psi\hat{r}}]}.
\label{eq:Btoriginal}
\end{equation}
In a spherical coordinate system\footnote{We use a right-handed spherical coordinate system where a positive radial field component points out of the star, the $\theta$ component is positive pointing from North to South and the $\phi$ component is positive in the clockwise direction as viewed from the South pole.} the scalars $\Phi$ and $\Psi$ take the form:
\begin{equation}
\Phi=S(r)c_{lm}P_{lm}e^{im\phi},
\label{eq:Phi}
\end{equation}
\begin{equation}
\Psi=T(r)c_{lm}P_{lm}e^{im\phi}. 
\label{eq:Psi}
\end{equation}
$P_{lm}\equiv P_{lm}(\cos{\theta})$ is the associated Legendre polynomial of mode $l$ and order $m$ and
\begin{equation}
c_{lm}\equiv\sqrt{\frac{2l+1}{4\pi}\frac{(l-m)!}{(l+m)!}} 
\label{eq:clm}
\end{equation}
is a normalisation constant. $S(r)$ and $T(r)$ are functions describing the radial behaviour of the magnetic field components. Determining the complete form of these functions from observations of stellar magnetic fields is impossible, however values can be obtained at stellar surfaces ($r=R_\star$).

\begin{figure}\centering
	\includegraphics[width=\columnwidth]{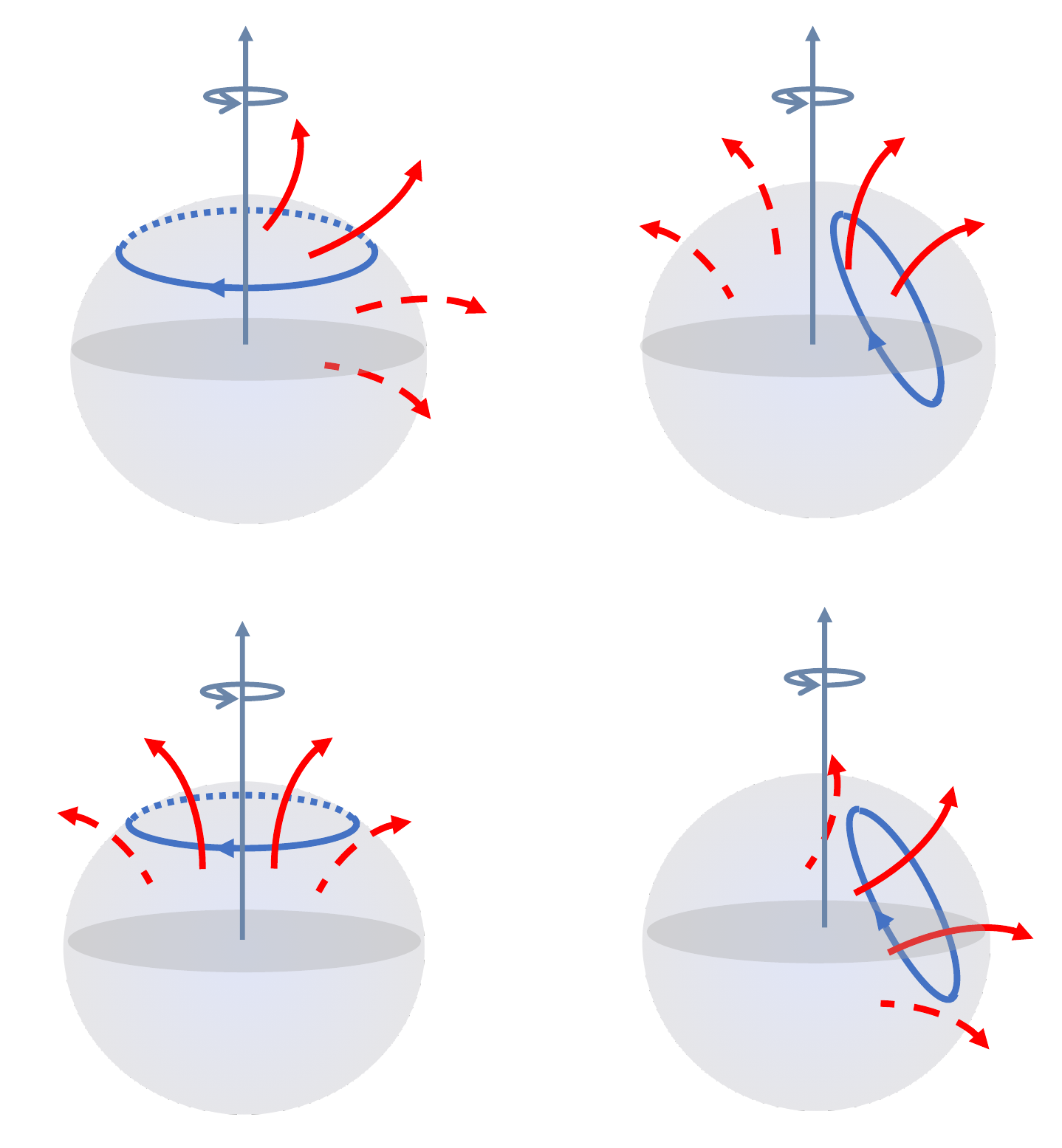}
	\caption{The cartoon shows four different combinations of symmetries (axisymmetric and non-axisymmetric relative to the rotation axis) of the poloidal (red arrows) and toroidal (blue arrows) fields. Poloidal field lines that link with the toroidal field are represented by solid lines, the ones that do not are dashed.} 
	\label{fig:cartoon}
\end{figure}

The Zeeman-Doppler imaging technique \citep{Semel1989} describes the large-scale (low $l$ modes) magnetic fields at the surfaces of stars in terms of $\alpha_{lm}$, $\beta_{lm}$, and $\gamma_{lm}$ coefficients \citep[e.g.][]{Donati2006,Vidotto2016}:
\begin{equation}
\begin{split}
\boldsymbol{B}_{\rm{pol}}(\theta,\phi)&=\sum_{lm}\alpha_{lm}c_{lm}P_{lm}e^{im\phi}\hat{r}\\
& +\sum_{lm}\frac{\beta_{lm}}{(l+1)}c_{lm}\dv{P_{lm}}{\theta}e^{im\phi}\hat{\theta}\\
&+\sum_{lm}\frac{\beta_{lm}im}{(l+1)\sin\theta}c_{lm}P_{lm}e^{im\phi}\hat{\phi},
\label{eq:sBp} 
\end{split}
\end{equation}
\begin{equation}
\begin{split}
\boldsymbol{B}_{\rm{tor}}(\theta,\phi)&=\sum_{lm}\frac{\gamma_{lm}im}{(l+1)\sin\theta}c_{lm}P_{lm}e^{im\phi}\hat{\theta}\\
&-\sum_{lm}\frac{\gamma_{lm}}{(l+1)}c_{lm}\dv{P_{lm}}{\theta}e^{im\phi}\hat{\phi}.
\label{eq:sBt} 
\end{split}
\end{equation}
These expressions are consistent with the general form of the poloidal and toroidal fields evaluated at the stellar surface when
\begin{equation}
S(R_\star)=\frac{\alpha_{lm}R_\star^2}{l(l+1)},\quad {\dv{S(r)}{r}}\Big\rvert_{r=R_\star}=\frac{\beta_{lm}R_\star}{(l+1)}
\label{eq:S_rstar}
\end{equation}
and 
\begin{equation}
T(R_\star)=\frac{\gamma_{lm}R_\star}{(l+1)}.
\label{eq:T_rstar}
\end{equation}
Given the surface magnetic field, magnetic energies are estimated by calculating the mean squared magnetic flux density $\langle{B^2}\rangle$. For instance, in the case of the poloidal energy\footnote{When calculating the mean squared magnetic flux density we integrate over a full sphere, dividing by the solid angle of $\Omega=$ 4$\pi$.}: $\langle{B^2}_{\rm{pol}}\rangle=\frac{1}{\Omega}\int \boldsymbol{B}_{\rm{pol}}\cdot\boldsymbol{B}_{\rm{pol}}\mathrm{d}\Omega$. Accordingly, the fraction of axisymmetric poloidal magnetic field energy is given by\footnote{We define axisymmetric as $m=0$.}: $\langle{B^2_{\rm{pol},m=0}}\rangle/\langle{B^2}_{\rm{pol}}\rangle$. The toroidal energy and axisymmetry fraction are calculated analogously.

\subsection{Magnetic helicity density}

Magnetic helicity can be defined as $H=\int \boldsymbol{A}\cdot \boldsymbol{B} \,\rm{dV}$ \citep{Woltjer1958}, where $\boldsymbol{A}$ is a vector potential corresponding to the magnetic field $\boldsymbol{B}$. As magnetic helicity is a quantity measuring the linkage of fields within a \textit{volume}, our \textit{surface} magnetic fields limit us to evaluating the magnetic helicity density $h=\boldsymbol{A}\cdot \boldsymbol{B}$. The separation of the magnetic field into its poloidal and toroidal components is particularly useful in this regard. It dispenses with the need to invoke a gauge \citep{Berger2018} since the usual gauge field (the corresponding potential field with the same boundary flux) has zero helicity. In addition, in a spherical coordinate system, toroidal field lines lie purely on spherical surfaces while poloidal field lines pass through these surfaces. This makes visualising the linkage of field lines straightforward. Fig. \ref{fig:cartoon} illustrates how poloidal fields lines (shown red) that pass through the stellar surface may link through loops of toroidal field (shown blue) that lie on the stellar surface. It is notable that in these examples, only some fraction of the poloidal field links with the toroidal field line shown.

Interpreting magnetic helicity as the linking of poloidal and toroidal fields \citep{Berger1985,Berger2018} allows the helicity density to be calculated for any stellar magnetic map given only its $\alpha_{lm}$ and $\gamma_{lm}$ coefficients and the stellar radius $R_\star$ \citep{Lund2020}: 
\begin{equation}
\begin{split}
    h(\theta,\phi)&=\Re\bigg(\sum_{lm}\sum_{l^\prime m^\prime}\frac{\alpha_{lm}\gamma_{l^\prime m^\prime}R_\star}{(l^\prime+1)l(l+1)}c_{lm}c_{l^\prime m^\prime}e^{i\phi(m+m^\prime)}\\
    &\bigg(P_{lm}P_{l^\prime m^\prime}\bigg(l(l+1)-\frac{mm^\prime}{\sin^2\theta}\bigg)+\dv{P_{lm}}{\theta}\dv{P_{l^\prime m^\prime}}{\theta}\bigg)\bigg).
\end{split}
\label{eq:surfaceh}
\end{equation}
The magnetic helicity density, as expressed in Eq. \ref{eq:surfaceh}, depends on the $\alpha_{lm}$ and $\gamma_{lm}$ coefficients, but not the $\beta_{lm}$ coefficients found in the $\theta$ and $\phi$ components of the poloidal field. This is because only the radial component of the poloidal field ($\boldsymbol{B}_{\rm{pol},r}$) passes through the spherical surfaces containing the toroidal field.

When comparing different magnetic maps, e.g. for different stars, it is often useful to summarise the overall helicity with a single number. For this purpose, we consider an average value across the hemisphere facing the observer. We note that typically only one hemisphere is fully visible as part of the star never comes into view. Furthermore, we take the absolute value of the averaged helicity density as we are interested in comparing magnitudes, not signs. This absolute average helicity density (|$\langle{h}\rangle$|) will for simplicity be referred to as the ``helicity density'' for the remainder of this paper. 

\begin{figure}\centering
	\includegraphics[width=\columnwidth]{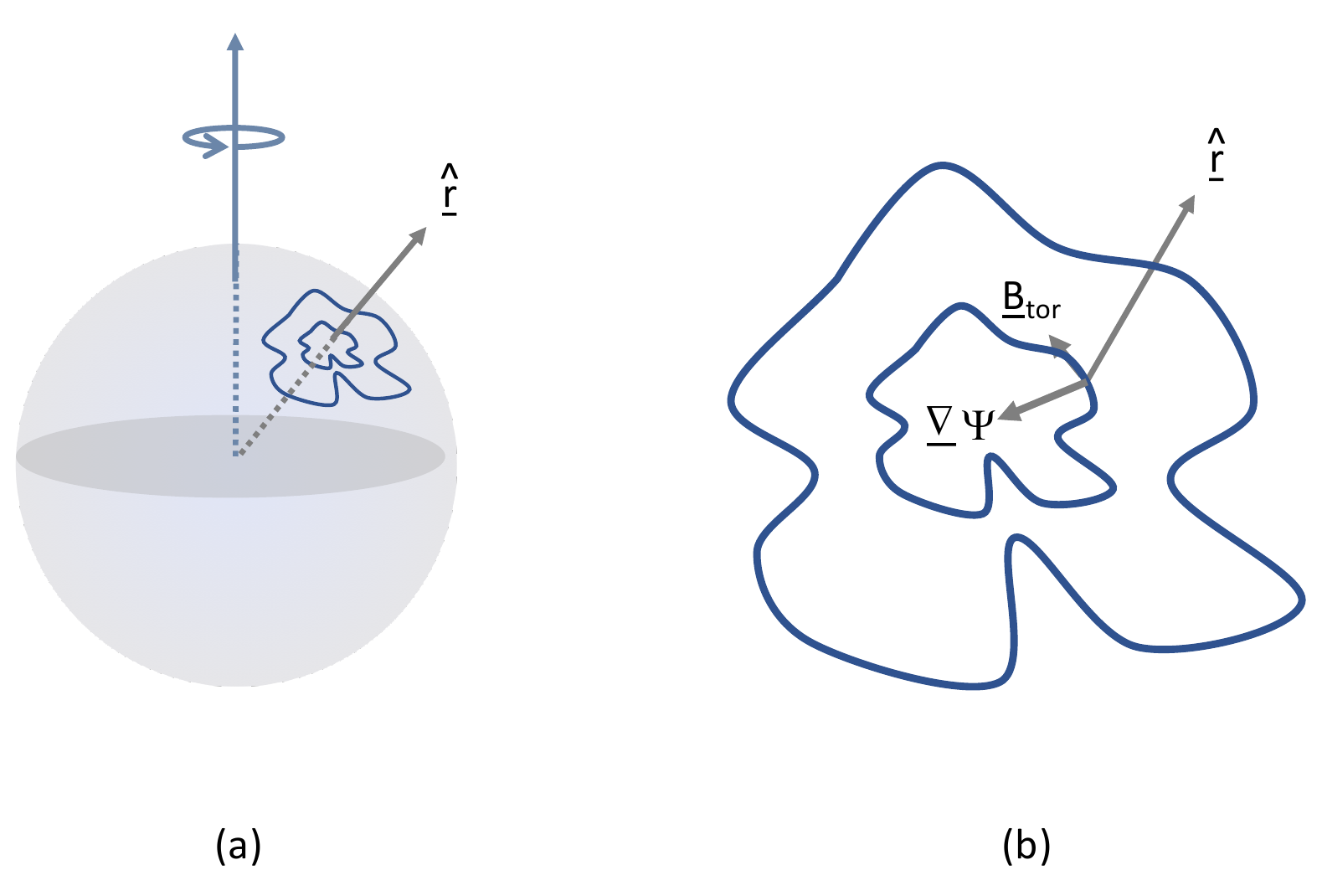}
	\caption{(a) Contours of the function $\Psi$ lying on the surface of the star are shown in blue. (b) This shows an enlarged version, illustrating that $\boldsymbol{B}_{\rm{tor}} =\grad{\Psi}\cross\hat{r}$.} 
	\label{fig:cartoon_contours}
\end{figure}

 It is possible to visualise the linkage of poloidal and toroidal fields that results in helicity density at the stellar surface through maps showing the strength of $\boldsymbol{B}_{\rm{pol},r}$ (calculated from the $\alpha_{lm}$ according to Eq. \ref{eq:sBp}) with the field lines of $\boldsymbol{B}_{\rm{tor}}$ superimposed. Expanding Eq. \ref{eq:Btoriginal} as
\begin{equation}
\boldsymbol{B}_{\rm{tor}}=\grad{\Psi}\cross\hat{r}
\label{eq:Btor_deriv}
\end{equation}
shows that the contours of $\Psi$ correspond to the field lines of $\boldsymbol{B}_{\rm{tor}}$ (see Fig. \ref{fig:cartoon_contours}). In particular, at the stellar surface,
\begin{equation}
    \Psi=\frac{\gamma_{lm}R_\star}{(l+1)}c_{lm}P_{lm}e^{im\phi},
    \label{eq:psi_surface}
\end{equation}  
from Eqs. \ref{eq:Psi} and \ref{eq:T_rstar}.
\section{Stellar sample}
\label{sec:sample}
    
\begin{figure}\centering
	\includegraphics[width=\columnwidth]{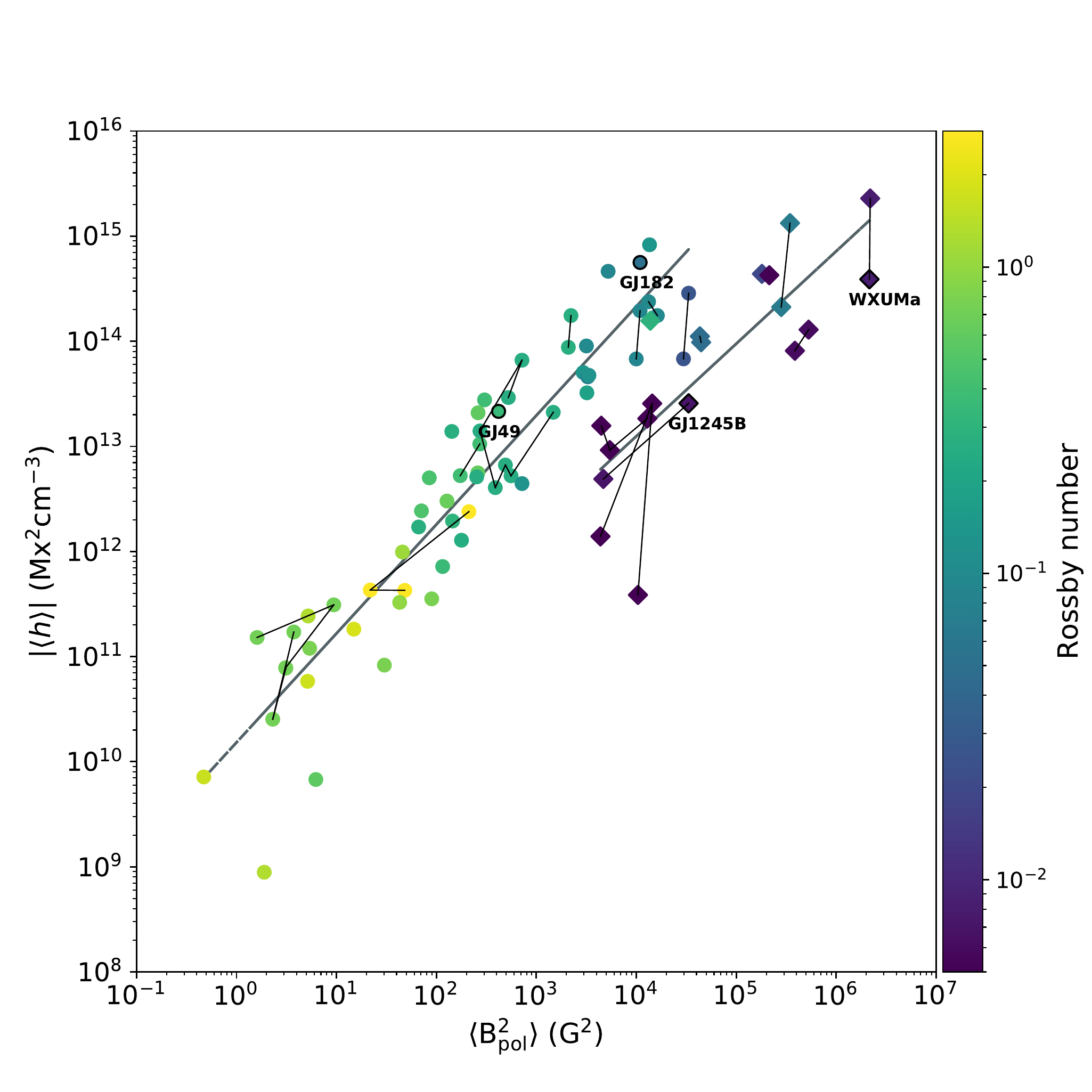}
	\caption{Absolute helicity density averaged across a single hemisphere versus the mean squared poloidal magnetic flux density ($l\leq$ 4). The colour of the symbols correspond to Rossby number and the shape of the symbols splits the sample into two mass groups; circles represent $M_\star$ > 0.5 M$_\odot$ and diamonds represent $M_\star\leq$ 0.5 M$_\odot$. When there are multiple measurements for the same star these are connected by lines. The thick grey lines show the best fit of \textbf{|$\langle{h\,}\rangle$| = $\langle{B^2_{\rm{pol}}\,}\rangle^\alpha 10^\beta$} for stars in the two mass groups; $\alpha=1.04\,\pm\,0.05$\textbf{, $\beta=10.18\,\pm\,0.13$} for $M_\star$ > 0.5 M$_\odot$ (circles) and $\alpha=0.88\,\pm\,0.15$\textbf{, $\beta=9.57\,\pm\,0.74$} for $M_\star\leq$ 0.5 M$_\odot$ (diamonds). The stars outlined are shown in Fig. \ref{fig:maps}.}
	\label{fig:slope}
\end{figure}

Our sample of stellar magnetic maps are all created using Zeeman-Doppler imaging. They describe the magnetic fields of 51 different stars, 15 of which are represented by multiple maps. The stars range in spectral type from F to M, and in mass from 0.1-1.34 M$_\odot$. Details of each star/map are provided in Table \ref{table:StellarSample}, along with the calculated helicity densities and magnetic energy components. For the sake of a fair comparison every magnetic map is evaluated to the same resolution, which means every calculation is performed up to the same $l$-mode ($l\leq 4$) even when higher modes are available \citep{Lund2020}.    

\section{The role of axisymmetry in helicity density}
\label{sec:results}

\begin{figure*}\centering
	\includegraphics[width=\textwidth]{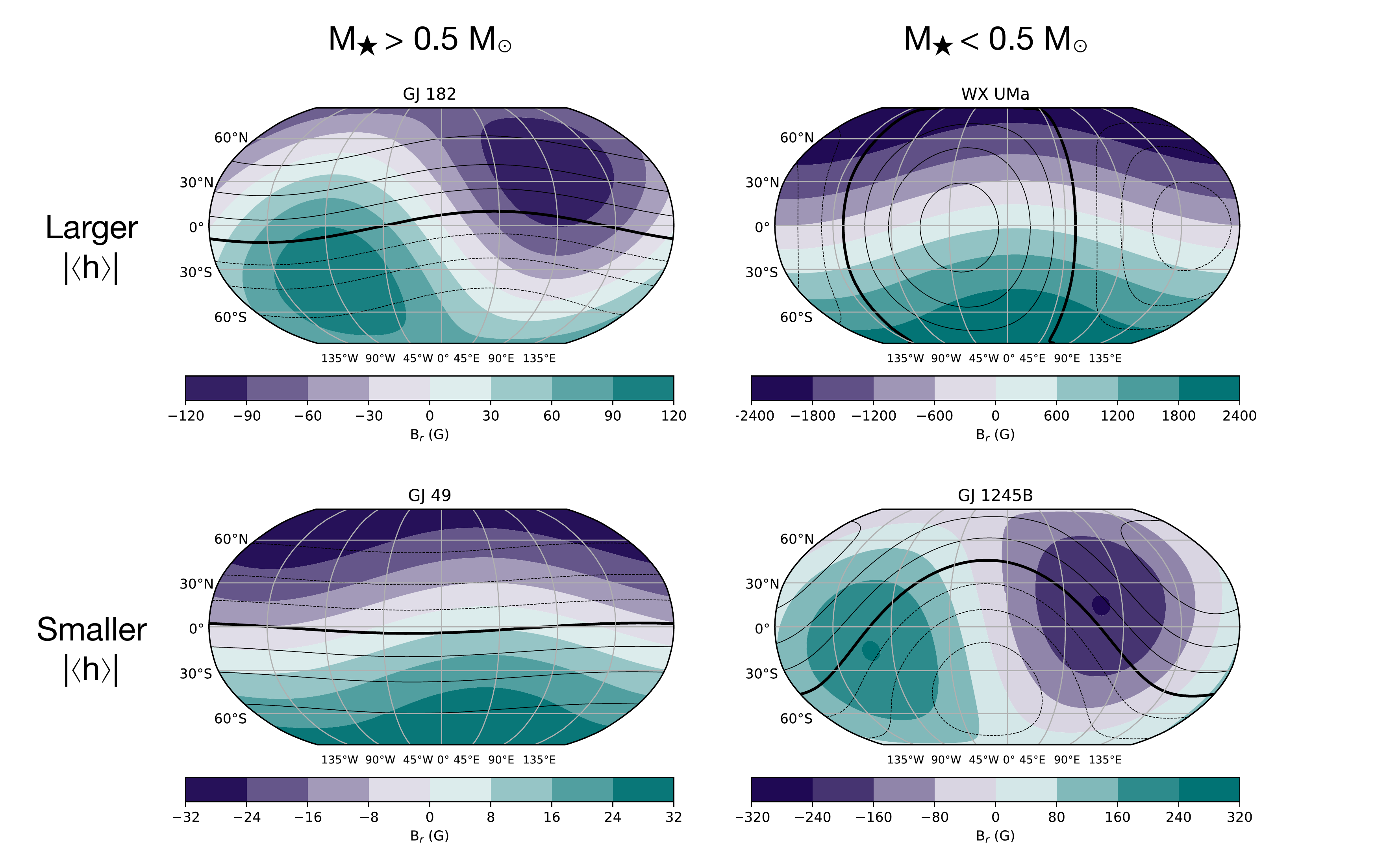}
	
	\caption{A visualisation of the linkage of the dipole ($l=1$) poloidal and toroidal field components of GJ 182, WX UMa (2008), GJ 49 and GJ 1245B (2006). The colour shows the strength of the radial magnetic (poloidal) field, and the black contours represent the toroidal magnetic field lines. The heavy black contour separates regions of positive (solid) and negative (dashed) toroidal field. These examples correspond roughly to the four classes shown in Fig. \ref{fig:cartoon}.}
	\label{fig:maps}
\end{figure*}

The very well-defined dependence of helicity density on the toroidal field of $|\langle{h\,}\rangle|$ $\propto$ $\langle{\rm{B_{tor}}^2_{}\,\rangle}^{0.86\,\pm\,0.04}$ was shown in \citet{Lund2020}. One enduring puzzle, however, is that the dependence on the poloidal field revealed two branches, as shown in Fig. \ref{fig:slope}.  The higher-mass branch ($M_\star>$ 0.5 M$_\odot$) follows |$\langle{h\,}\rangle$| $=\langle{B^2_{\rm{pol}}\,}\rangle^{1.04\,\pm\,0.05}10^{10.18\,\pm\,0.13}$ and the lower-mass branch ($M_\star\leq$ 0.5 M$_\odot$) follows |$\langle{h\,}\rangle$| $=\langle{B^2_{\rm{pol}}\,}\rangle^{0.88\,\pm\,0.15}10^{9.57\,\pm\,0.74}$. When fitting these power laws the sample is split specifically at 0.5 M$_\odot$ because a number of magnetic properties, including helicity density, have been shown to change behaviour across this value \citep{Donati2008,Morin2008b,Morin2010,See2015,Lund2020}. Fig. \ref{fig:slope} shows that the lowest mass stars have higher poloidal energies than higher-mass stars with the same helicity density. The lowest mass stars in our sample also typically have the lowest Rossby numbers, as indicated by the colours in the plot.

It appears from Fig. \ref{fig:slope} that the lower-mass fully-convective stars have excess poloidal field that does not contribute to their helicity density. In order to explore the distribution of the poloidal and toroidal fields on these two branches and to determine their contribution to the helicity density, we plot maps showing how their poloidal and toroidal fields link. As an example, Fig. \ref{fig:maps} presents maps for GJ 182, WX UMa, GJ 49 and GJ 1245B. These stars are highlighted in Fig. \ref{fig:slope} and represent two pairs of stars with approximately the same helicity density, and thus similar toroidal energies. Each pair consists of one star from the higher-mass branch and one star from the lower-mass branch, and the two pairs are distinguished by the magnitude of their helicity densities (GJ 182 and WX UMa (top row) have higher helicity densities than GJ 49 and GJ 1245B). To illustrate field linkages at the largest scale the maps show the dipole ($l=1$) mode. We note that these four stars fall into the four categories shown in Fig. \ref{fig:cartoon}.

\begin{figure}\centering
	\includegraphics[width=\columnwidth]{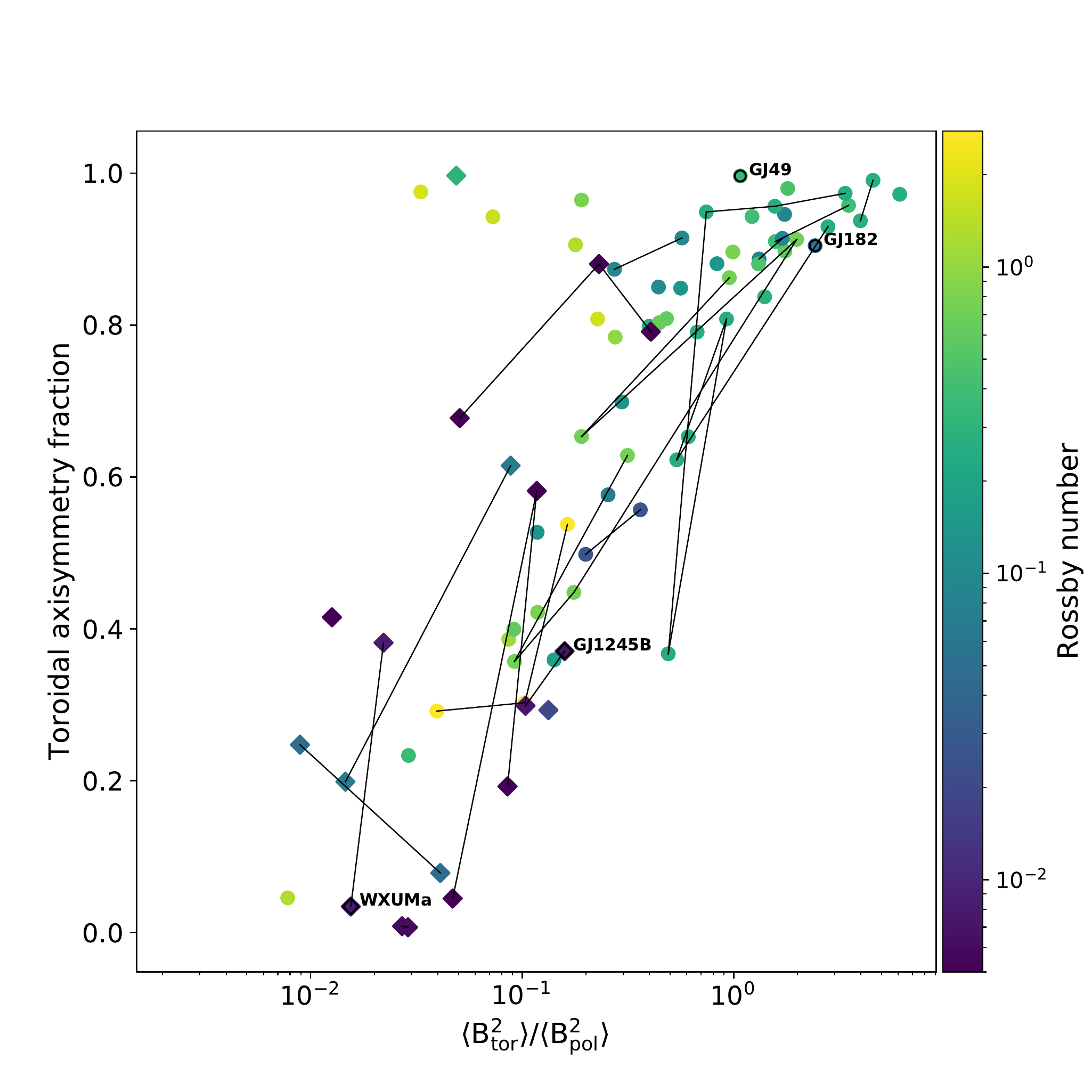}
	\caption{Axisymmetric toroidal energy as a fraction of total toroidal energy versus the ratio of the mean squared toroidal to poloidal magnetic flux densities ($l\leq$ 4). The symbols are the same as in Fig. \ref{fig:slope}.}
	\label{fig:excess}
\end{figure}

\begin{figure}\centering
	\includegraphics[width=\columnwidth]{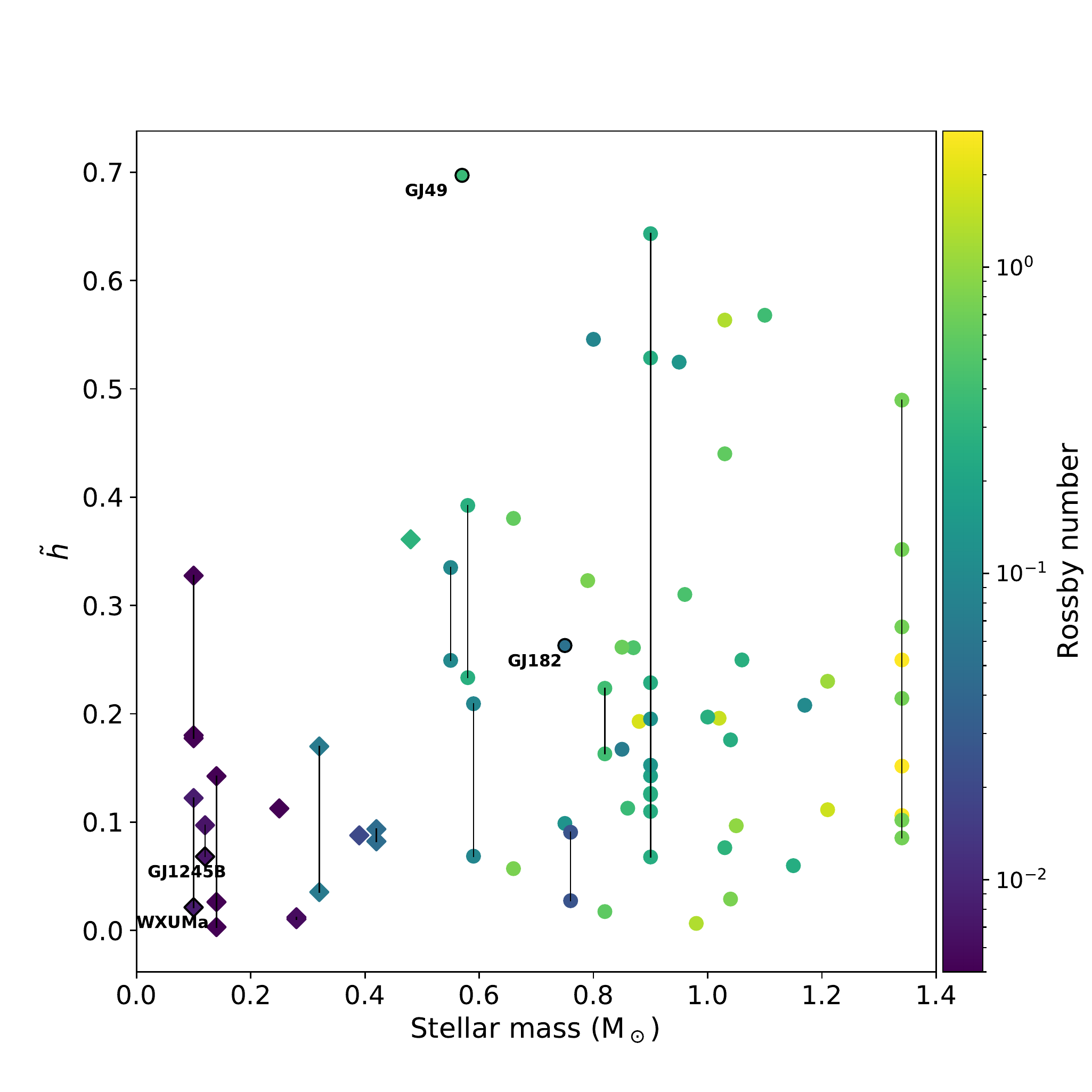}
	
	\caption{The helicity energy fraction ($l\leq 4$), defined as $\tilde{h}\equiv$ |$\langle{h\,}\rangle$|/$\langle{R_\star B^2\,}\rangle$, versus stellar mass. The symbols are the same as in Fig. \ref{fig:slope}. }
	\label{fig:h_normalised}
\end{figure}

\begin{figure}\centering
	\includegraphics[width=\columnwidth]{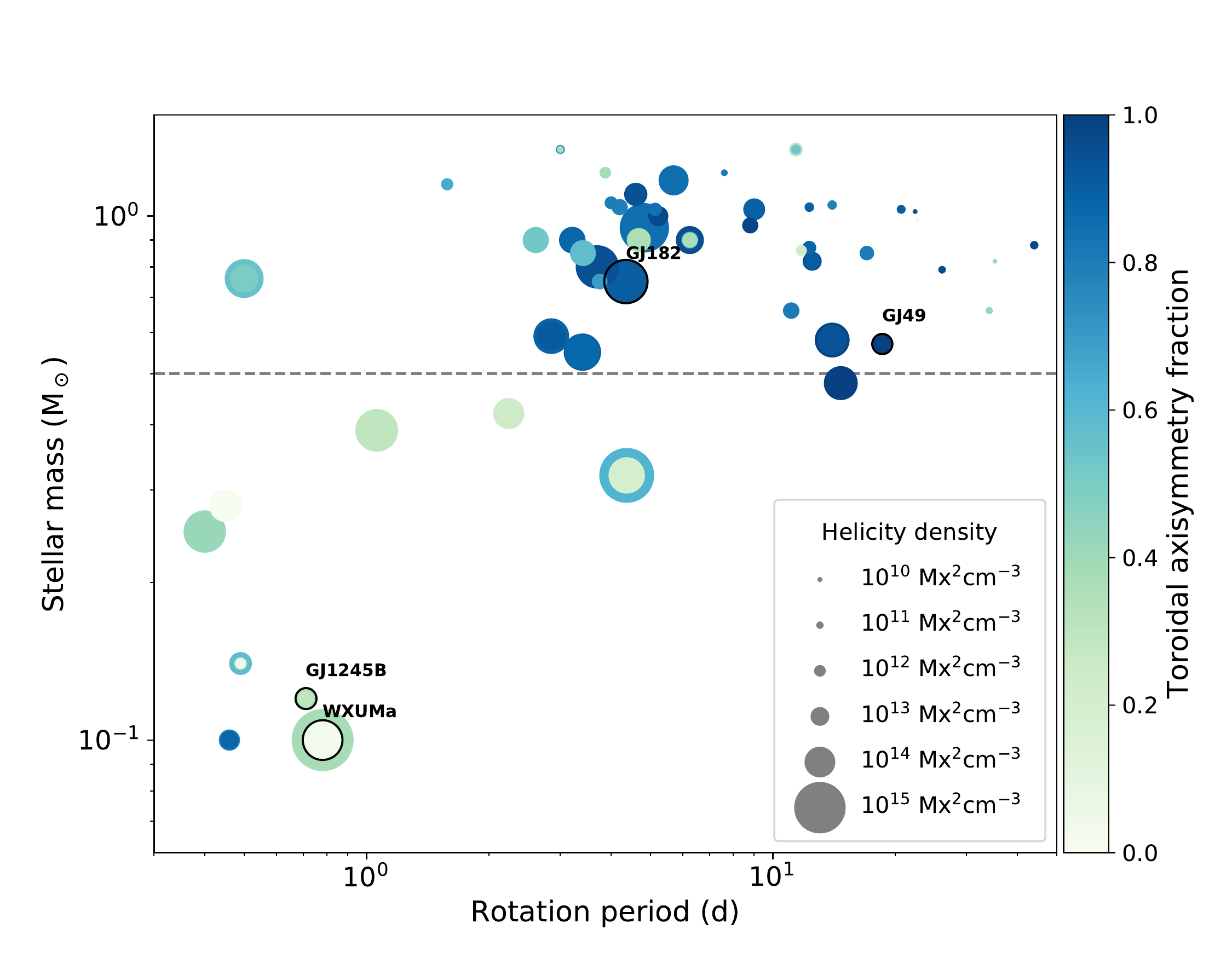}\\
	\includegraphics[width=\columnwidth]{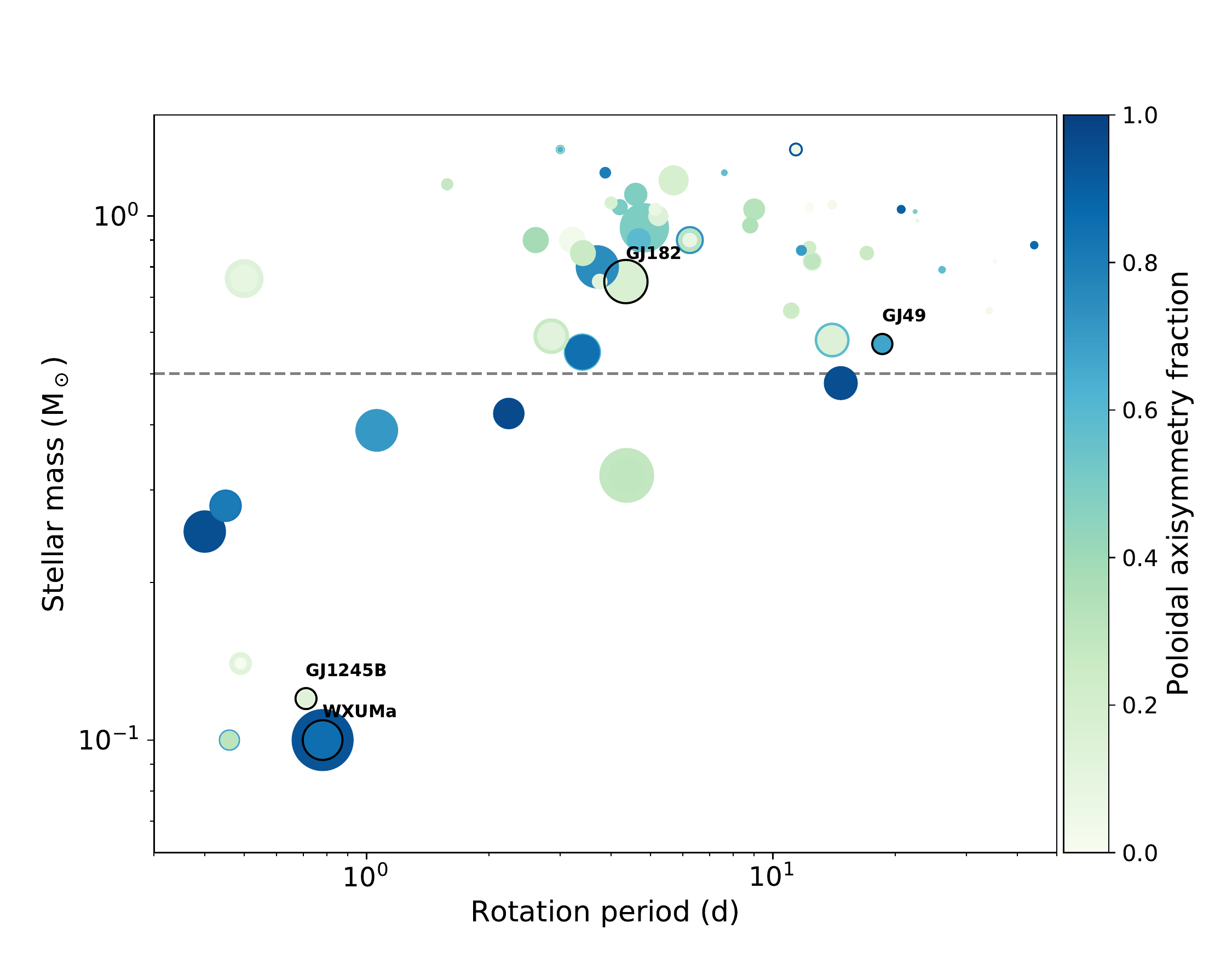}
	\caption{Helicity density of the large-scale magnetic fields ($l\leq 4$) of our stellar sample shown according to stellar mass and rotation period. The dashed lines split the sample at 0.5 M$_\odot$. The size of the symbol indicates relative strength of the helicity density. The colour corresponds to the fractional axisymmetry of the toroidal (\textit{top}) and poloidal (\textit{bottom}) magnetic energies. The stars outlined are shown in Fig. \ref{fig:maps}.}
	\label{fig:confusagram}
\end{figure}

The clearest trend to emerge is in the toroidal field. It is notable that the stars on the higher-mass branch, GJ 182 and GJ 49 (left column of Fig. \ref{fig:maps}), both have fairly axisymmetric toroidal fields, whilst the stars on the lower mass branch, WX UMa and GJ 1245B, have non-axisymmetric toroidal fields. We can quantify this trend by plotting the fraction of the toroidal field energy that is held in axisymmetric modes as a function of the ratio of toroidal to poloidal energy (see Fig. \ref{fig:excess}). Lower-mass stars tend to have toroidal fields that are non-axisymmetric and magnetic energy budgets that are dominated by the poloidal field.

\citet{Lund2020} showed that stars with a similar helicity density have toroidal fields of similar strengths. As can be seen from Fig. \ref{fig:maps}, however, these toroidal fields may have different symmetries if the stars lie on two different branches. The fact that these different symmetries are able to produce the same helicity density is because the poloidal field, although typically stronger on the lower-mass branch, can also be of different symmetry to the toroidal field. In both rows of Fig. \ref{fig:maps}, the transition from the higher-mass branch to the lower-mass branch is accompanied by an increase in the strength of the poloidal field. At the same time, for the lower mass stars, toroidal field lines enclose regions of both positive and negative poloidal field such that the total linked flux is relatively small. The difference in magnetic field topology roughly cancels the difference in field strength, such that the helicity is very similar within each pair, although it is arrived at very differently.

To separate the effects of field strength and geometry Fig. \ref{fig:h_normalised} plots stellar mass against the ``helicity energy fraction''; $\tilde{h}\equiv$ |$\langle{h\,}\rangle$|/$\langle{R_\star B^2\,}\rangle$. Dividing the helicity density by the mean squared magnetic flux density and stellar radius results in a dimensionless helicity density normalised by magnetic field strength. Fig. \ref{fig:h_normalised} shows an even spread of low helicity energy fractions across the entire range of stellar masses, however only the higher mass stars ($M_\star>$ 0.5 M$_\odot$) exceed a fraction of 0.35. This illustrates that even though the lower-mass stars are among those whose dynamos are most efficient at injecting magnetic energy into the largest spatial scales that ZDI is able to detect and map \citep[e.g.][]{Morin2008b}, they are apparently \textit{less} efficient at generating helicity at these largest scales. This is most likely because of the inefficient linking between their poloidal and toroidal magnetic field components.

Inefficient linking can be arrived at in more than one way.  For the higher-helicity pair in Fig. \ref{fig:maps} (GJ 182, WX UMa), the lower-mass star has a strongly axisymmetric poloidal field, and since the toroidal field is non-axisymmetric, this combination produces inefficient linking. In the case of the lower-helicity pair (GJ 49 and GJ 1245B), the lower-mass star’s poloidal and toroidal fields are both strongly non-axisymmetric; nonetheless, they still offset each another by approximately 90 degrees, which again gives inefficient linking.

We can place these trends in a broader context by showing how the axisymmetry and helicity density varies across the stellar mass-rotation period plane. This is shown in Fig. \ref{fig:confusagram} which also shows separately the variation of the toroidal and poloidal axisymmetry fractions. High-helicity stars exist in both mass ranges. The decline in the axisymmetry of the toroidal fields with decreasing mass is apparent, but the trends in poloidal axisymmetry are more complex. The reasons for this, and the potential role of magnetic cycles, are not yet clear.

Fig. \ref{fig:confusagram} also demonstrates very clearly the differences in the strength and structure of the magnetic field that is possible for stars in the bimodal regime, which includes the stars of lowest mass and shortest period. Two of our example stars (WX Uma and GJ 1245B) lie in this regime. They have similar masses (0.1 and 0.12 M$_\odot$ respectively) and rotation periods (0.78 and 0.71 days, respectively) but their magnetic fields and the helicity densities they support are quite different.

\section{Discussion and conclusions}
\label{sec:conclusion}

Helicity measures the linkage within a field. By studying the linkage of the poloidal and toroidal components of stellar magnetic fields, we can learn about the underlying dynamo processes generating the field, and thereby the form of the magnetic cycles that might take place and the evolution of stellar fields as stars spin down over their main sequence lifetimes. This study reveals that stars in different mass ranges, which may be either fully or partially convective, generate their helicity through different forms of toroidal/poloidal field linkage. 

For partially convective stars (those that lie on the higher-mass branch in Fig. \ref{fig:slope}) the toroidal fields are mainly axisymmetric. An increase in rotation rate (or a decrease in Rossby number) generally leads to increased helicity density. For fully-convective stars (those that lie on the lower-mass branch in Fig. \ref{fig:slope}) the toroidal fields are mainly non-axisymmetric. For the lowest-mass stars, a bimodal behaviour is present in the form and strength of the magnetic field \citep{Morin2011} which may be weak with a non-axisymmetric poloidal field, or strong with an axisymmetric poloidal field. These two types correspond to low and high helicity densities respectively.  This may have implications for the possibility that this represents a bimodality in the dynamo operating in this regime \citep{Morin2011}.

Stars can evolve from one mass-branch to the other if their field structure changes, for instance as a result of their internal structure changing from mainly convective to mainly radiative at a very young age, or if they transition from one bimodal dynamo mode to the other. Furthermore, stars can evolve along each branch as their rotation rates decay with age. Their field linkages can also evolve on much shorter timescales due to magnetic cycles. It is notable in Fig. \ref{fig:excess} that where there are multiple observations of a star, taken at different times, these typically follow the trend that an increase in the ratio of $B^2_{\rm tor}$ to $B^2_{\rm pol}$ leads to an increase in the axisymmetry of the toroidal field. A similar behaviour is seen in Fig. \ref{fig:slope} where for each star with multiple observations, these all lie within the scatter about the best-fit line.

It is not clear what causes the non-axisymmetry of the toroidal fields in the lowest-mass stars. Their deep-seated convection may produce bipoles that emerge through the stellar surface with  randomised axial tilts, leading to a lack of axisymmetry in the toroidal field. Their low surface differential rotation may also reduce the shearing of bipoles and hence the diffusive cancellation of poloidal field that results. Both of these processes, however, occur at length scales well below what can be resolved by these Zeeman-Doppler field measurements. The field characteristics that are most robustly recovered by Zeeman-Doppler imaging (field axisymmetry and the ratio of poloidal to toroidal field \citep{Lehmann2019}) are nonetheless the very ones that underpin the helicity density.

In summary, we find that lowest-mass stars tend to be inefficient at generating helicity on the largest scales, given their magnetic energy. The helicity density at a stellar surface depends not only on the stellar radius and the strengths of the individual poloidal and toroidal field components (see Eq. \ref{eq:surfaceh}) but also on their spatial distribution {\it relative to each other}. The fraction of the poloidal flux that links with the toroidal flux is maximised when the axes of symmetry of the two fields align perfectly. The bottom row of Figure \ref{fig:cartoon} illustrates two ways in which such an alignment is possible: where both fields are axisymmetric (left) and where both are non-axisymmetric (right). Conversely, if the poloidal and toroidal axes of symmetry are orthogonal to each other there is no linkage, and consequently no helicity. Different orientations with different amounts of field linking can nonetheless result in the same helicity density due to the dependence on field strengths. In short, to achieve a full understanding of the source of the helicity density at a stellar surface, it is not enough simply to look at, for example, the radius and field strength, it is also necessary to produce a surface map showing the field linkage. It is only through a combination of all these components that a clear picture can be formed.

\newcolumntype{Y}{>{\centering\arraybackslash}X}
\newcolumntype{s}{>{\hsize=.6\hsize}Y}
\begin{center}
\begin{table*}
\centering
\caption{Our stellar sample, with the four stars we focus on in this paper highlighted in bold (GJ 182, GJ 49, GJ 1245B and WX UMa). From left to right the columns show: star name, mass, radius, rotation period,Rossby number,  absolute helicity density averaged across the visible hemisphere, $\langle{B^2_{\rm{pol}}\,}\rangle$, poloidal axisymmetric magnetic energy as a fraction of poloidal energy, $\langle{B^2_{\rm{tor}}\,}\rangle$, toroidal axisymmetric magnetic energy as a fraction of toroidal energy, $l_{\rm{max}}$ and observation epoch. The helicity density and the energies are all calculated for $l\leq 4$. References for the stellar parameters are given in the last column, where references to the papers where the magnetic maps were published are in italic. A more comprehensive table of parameters for these stars can be found in \citet{Vidotto2014}.}
\begin{tabularx}{\textwidth}{ l s s s s Y Y s Y s s Y r}
\hline
        Star ID&$M_{\star}$&$R_{\star}$&$P_{\rm{rot}}$&$R_o$&|$\langle{h\,}\rangle$|&$\langle{B^2_{\rm{pol}}\,}\rangle$&Pol&$\langle{B^2_{\rm{tor}}\,}\rangle$&Tor&$l_{\rm{max}}$&Obs.&Ref.\T\\ 
    &(M$_{\odot}$)&(R$_{\odot}$)&(d)&&(Mx$^2\rm{cm}^{-3}$)&(G$^2$)&Axi&(G$^2$)&Axi&&epoch&\B \\
    \hline
\multicolumn{13}{l}{\textbf{Solar-like stars}\T} \\
HD 3651	&	0.88	&	0.88	&	44.0	&	1.916	&	1.82E+11	&	1.49E+01	&	0.87	&	4.96E-01	&	0.98	&	10	&	--	&	1, 2, \textit{3}	\\
HD 9986	&	1.02	&	1.04	&	22.4	&	1.621	&	7.16E+09	&	4.71E-01	&	0.50	&	3.43E-02	&	0.94	&	10	&	--	&	1, 2, \textit{3}	\\
HD 10476	&	0.82	&	0.82	&	35.2	&	0.576	&	6.77E+09	&	6.23E+00	&	0.00	&	5.69E-01	&	0.40	&	10	&	--	&	1, 4, 2, \textit{3}	\\
HD 20630&1.03	&	0.95&9.00&0.593	& 2.09E+13&2.61E+02&0.32&4.56E+02&0.90&10&Oct 2012&1, 5, 2, \textit{6}\\		
HD 22049	&	0.86	&	0.77	&	11.76&0.366	& 7.19E+11&1.16E+02&0.70&3.35E+00&0.23&10&-&1, 5, 2, \textit{7}\\																						
HD 39587	&	1.03	&	1.05	&	5.136	&	0.295	&	1.95E+12	&	1.45E+02	&	0.07	&	2.04E+02	&	0.84	&	10	&	--	&	1, 5, 2, \textit{3}	\\
HD 56124	&	1.03	&	1.01	&	20.7	&	1.307	&	2.43E+11	&	5.22E+00	&	0.90	&	9.32E-01	&	0.91	&	10	&	--	&	1, 2, \textit{3}	\\
HD 72905	&	1	&	1	&	5.227	&	0.272	&	1.39E+13	&	1.43E+02	&	0.15	&	8.70E+02	&	0.97	&	10	&	--	&	1, 5, 2, \textit{3}	\\
HD 73350	&	1.04	&	0.98	&	12.3	&	0.777	&	3.54E+11	&	8.99E+01	&	0.00	&	8.91E+01	&	0.90	&	10	&	--	&	1, 8, 2, \textit{3}	\\
HD 75332	&	1.21	&	1.24	&	3.870	&	>1.105	&	9.88E+11	&	4.58E+01	&	0.80	&	3.96E+00	&	0.39	&	15	&	--	&	1, 5, 2, \textit{3}	\\
HD 78366	&	1.34	&	1.03	&	11.4	&	>2.781	&	2.39E+12	&	2.11E+02	&	0.94	&	8.36E+00	&	0.29	&	10	&	2008	&	1, 2, \textit{9}	\\
…	&	…	&	…	&	…	&	…	&	4.27E+11	&	4.83E+01	&	0.06	&	7.91E+00	&	0.54	&	…	&	2010	&	…	\\
…	&	…	&	…	&	…	&	…	&	4.29E+11	&	2.18E+01	&	0.77	&	2.25E+00	&	0.30	&	…	&	2011	&	…	\\
HD 101501	&	0.85	&	0.9	&	17.04	&	0.663	&	3.02E+12	&	1.28E+02	&	0.26	&	5.68E+01	&	0.80	&	10	&	--	&	1, 5, 2, \textit{3}	\\
HD 131156A	&	0.90	&	0.80	&	6.25	&	0.256	&	2.11E+13	&	1.48E+03	&	0.30	&	4.12E+03	&	0.93	&	10	&	Aug 2007	&	10, 2, \textit{11}	\\
…	&	…	&	…	&	…	&	…	&	5.24E+12	&	5.58E+02	&	0.59	&	3.00E+02	&	0.62	&	…	&	Feb 2008	&	…	\\
…	&	…	&	…	&	…	&	…	&	6.64E+12	&	4.91E+02	&	0.07	&	4.54E+02	&	0.81	&	…	&	Jun 2009	&	…	\\
…	&	…	&	…	&	…	&	…	&	4.05E+12	&	3.89E+02	&	0.08	&	1.91E+02	&	0.37	&	…	&	Jan 2010	&	…	\\
…	&	…	&	…	&	…	&	…	&	1.40E+13	&	2.73E+02	&	0.34	&	2.03E+02	&	0.95	&	…	&	Jun 2010	&	…	\\
…	&	…	&	…	&	…	&	…	&	6.60E+13	&	7.18E+02	&	0.73	&	1.12E+03	&	0.96	&	…	&	Aug 2010	&	…	\\
…	&	…	&	…	&	…	&	…	&	2.91E+13	&	5.25E+02	&	0.27	&	1.77E+03	&	0.97	&	…	&	Jan 2011	&	…	\\
HD 131156B	&	0.66	&	0.55	&	11.1	&	0.611	&	5.59E+12	&	2.60E+02	&	0.25	&	1.25E+02	&	0.81	&	10	&	--	&	10, 2, \textit{3}	\\
HD 146233	&	0.98	&	1.02	&	22.7	&	1.324	&	8.87E+08	&	1.90E+00	&	0.09	&	1.48E-02	&	0.05	&	10	&	Aug 2007	&	12, 2, \textit{8}	\\
HD 166435	&	1.04	&	0.99	&	4.2	&	0.259	&	5.12E+12	&	2.53E+02	&	0.50	&	1.70E+02	&	0.79	&	10	&	--	&	1, 2, \textit{3}	\\
HD 175726	&	1.06	&	1.06	&	4.0	&	0.272	&	1.71E+12	&	6.65E+01	&	0.18	&	2.65E+01	&	0.80	&	10	&	--	&	1, 13, 2, \textit{3}	\\
HD 190771	&	0.96	&	0.98	&	8.80	&	0.453	&	5.02E+12	&	8.50E+01	&	0.35	&	1.53E+02	&	0.98	&	10	&	2007	&	9, 2, \textit{8}	\\
HD 201091A	&	0.66	&	0.62	&	34.1	&	0.786	&	8.29E+10	&	3.02E+01	&	0.03	&	3.57E+00	&	0.42	&	10	&	--	&	1, 5, 2, \textit{14}	\\
HD 206860	&	1.1	&	1.04	&	4.6	&	0.388	&	2.77E+13	&	3.03E+02	&	0.49	&	3.71E+02	&	0.94	&	10	&	--	&	1, 2, \textit{15}	\\
\multicolumn{13}{l}{\textbf{Young suns}\T} \\
AB Dor 	&	0.76	&	1.00	&	0.5	&	0.026	&	2.87E+14	&	3.34E+04	&	0.14	&	1.21E+04	&	0.56	&	25	&	Dec 2001	&	16, 17, 2, \textit{18}	\\	
…	&	…	&	…	&	…	&	…	&	6.80E+13	&	2.97E+04	&	0.09	&	5.92E+03	&	0.50	&	…	&	Dec 2002	&	….	\\	[2pt]
BD-16351	&	0.9	&	0.88	&	3.21	&	0.14$^{+0.01}_{-0.02}$	&	5.04E+13	&	2.94E+03	&	0.04	&	2.45E+03	&	0.88	&	15	&	Sep 2012	&	\textit{19}	\\	[2pt]
HII 296	&	0.9	&	0.93	&	2.61	&	0.13$^{+0.01}_{-0.01}$	&	4.74E+13	&	3.36E+03	&	0.38	&	3.96E+02	&	0.53	&	15	&	Oct 2009	&	\textit{19}	\\	[2pt]
HII 739	&	1.15	&	1.07	&	1.58	&	0.25$^{+0.01}_{-0.08}$	&	1.28E+12	&	1.78E+02	&	0.28	&	1.09E+02	&	0.65	&	15	&	Oct 2009	&	\textit{19}	\\	[2pt]
HIP 12545	&	0.95	&	1.07	&	4.83	&	0.14$^{+0.02}_{-0.02}$	&	8.28E+14	&	1.36E+04	&	0.49	&	7.63E+03	&	0.85	&	15	&	Sep 2012	&	\textit{19}	\\	[2pt]
HIP 76768	&	0.80	&	0.85	&	3.70	&	0.09$^{+0.03}_{-0.02}$	&	4.63E+14	&	5.23E+03	&	0.75	&	9.10E+03	&	0.95	&	15	&	May 2013	&	\textit{19}	\\	[2pt]
TYC 0486-4943-1	&	0.75	&	0.69	&	3.75	&	0.13$^{+0.03}_{-0.03}$	&	4.42E+12	&	7.19E+02	&	0.13	&	2.13E+02	&	0.70	&	15	&	Jun 2013	&	\textit{19}	\\	[2pt]
TYC 5164-567-1	&	0.90	&	0.89	&	4.68	&	0.19$^{+0.04}_{-0.05}$	&	3.23E+13	&	3.21E+03	&	0.59	&	4.54E+02	&	0.36	&	15	&	Jun 2013	&	\textit{19}	\\	[2pt]
TYC 6349-0200-1	&	0.85	&	0.96	&	3.41	&	0.07$^{+0.01}_{-0.02}$	&	4.58E+13	&	3.27E+03	&	0.26	&	8.35E+02	&	0.58	&	15	&	Jun 2013	&	\textit{19}	\\	[2pt]
TYC 6878-0195-1	&	1.17	&	1.37	&	5.70	&	0.10$^{+0.04}_{-0.03}$	&	9.02E+13	&	3.17E+03	&	0.19	&	1.40E+03	&	0.85	&	15	&	Jun 2013	&	\textit{19}	\\

\multicolumn{12}{l}{\textbf{Hot Jupiter Hosts}\T} \\
$\tau$ Boo	&	1.34	&	1.42	&	3	&	>0.732	&	1.52E+11	&	1.61E+00	&	0.48	&	1.54E+00	&	0.86	&	5	&	Jun 2006	&	20, 21, 2, \textit{22}	\\

…	&	…	&	…	&	…	&	…	&	3.11E+11	&	9.42E+00	&	0.59	&	1.80E+00	&	0.65	&	8	&	Jun 2007	&	20, 21, 2, \textit{23}	\\
…	&	…	&	…	&	…	&	…	&	7.83E+10	&	3.10E+00	&	0.13	&	6.16E+00	&	0.91	&	…	&	Jan 2008	&	20, 2, \textit{21}	\\
…	&	…	&	…	&	…	&	…	&	7.76E+10	&	3.12E+00	&	0.30	&	5.48E-01	&	0.45	&	…	&	Jun 2008	&	20, 2, \textit{21}	\\
…	&	…	&	…	&	…	&	…	&	2.53E+10	&	2.31E+00	&	0.62	&	2.12E-01	&	0.36	&	…	&	Jul 2008	&	20, 2, \textit{21}	\\
…	&	…	&	…	&	…	&	…	&	1.71E+11	&	3.75E+00	&	0.56	&	1.18E+00	&	0.63	&	…	&	Jun 2009	&	20, 21, 2, \textit{24}	\\

HD 73256	&	1.05	&	0.89	&	14	&	0.962	&	3.28E+11	&	4.30E+01	&	0.03	&	1.18E+01	&	0.78	&	4	&	Jan 2008	&	25, 2, \textit{24}	\\

HD 102195	&	0.87	&	0.82	&	12.3	&	0.473	&	2.43E+12	&	7.09E+01	&	0.23	&	9.30E+01	&	0.88	&	4	&	Jan 2008	&	26, 27, 2, \textit{24}	\\

\end{tabularx}
\end{table*}
\addtocounter{table}{-1}
\begin{table*}
\centering
\caption{-continued}
\begin{tabularx}{\textwidth}{l s s s s Y Y s Y s s Y r}
\hline
        Star ID&$M_{\star}$&$R_{\star}$&$P_{\rm{rot}}$&$R_o$&|$\langle{h\,}\rangle$|&$\langle{B^2_{\rm{pol}}\,}\rangle$&Pol&$\langle{B^2_{\rm{tor}}\,}\rangle$&Tor&$l_{\rm{max}}$&Obs.&Ref.\T\\ 
    &(M$_{\odot}$)&(R$_{\odot}$)&(d)&&(Mx$^2\rm{cm}^{-3}$)&(G$^2$)&Axi&(G$^2$)&Axi&&epoch&\B \\
    \hline

HD 130322	&	0.79	&	0.83	&	26.1	&	0.782	&	1.20E+11	&	5.40E+00	&	0.58	&	1.03E+00	&	0.96	&	4	&	Jan 2008	&	20, 28, 29, 2, \textit{24}	\\
HD 179949 	&	1.21	&	1.19	&	7.6	&	>1.726	&	5.82E+10	&	5.15E+00	&	0.57	&	1.17E+00	&	0.81	&	6	&	Jun 2007	&	12, 2, \textit{30}	\\
HD 189733	&	0.82	&	0.76	&	12.5	&	0.403	&	5.26E+12	&	1.73E+02	&	0.30	&	2.73E+02	&	0.91	&	5	&	Jun 2007	&	31, 2, \textit{32}	\\
…	&	…	&	…	&	…	&	…	&	1.05E+13	&	2.72E+02	&	0.17	&	9.50E+02	&	0.96	&	5	&	Jul 2008	&	…	\\

\multicolumn{13}{l}{\textbf{M dwarf stars}\T} \\
GJ 569A	&	0.48	&	0.43	&	14.7	&	<0.288	&	1.57E+14	&	1.38E+04	&	0.96	&	6.75E+02	&	1.00	&	5	&	Jan 2008	&	2, \textit{33}	\\
DS Leo	&	0.58	&	0.52	&	14	&	<0.267	&	1.76E+14	&	2.22E+03	&	0.58	&	1.01E+04	&	0.99	&	5	&	Jan 2007	&	2, \textit{33}	\\
…	&	…	&	…	&	…	&	…	&	8.76E+13	&	2.09E+03	&	0.15	&	8.31E+03	&	0.94	&	…	&	Dec 2007	&	…	\\

\textbf{GJ 182}	&	\textbf{0.75}	&	\textbf{0.82}	&	\textbf{4.35}	&	\textbf{0.054}	&	\textbf{5.62E+14}	&	\textbf{1.09E+04}	&	\textbf{0.17}	&	\textbf{2.65E+04}	&	\textbf{0.90}	&	\textbf{8}	&	\textbf{Jan 2007}	&	\textbf{2, \textit{33}}	\\
\textbf{GJ 49}	&	\textbf{0.57}	&	\textbf{0.51}	&	\textbf{18.6}	&	\textbf{<0.352}	&	\textbf{2.15E+13}	&	\textbf{4.19E+02}	&	\textbf{0.67}	&	\textbf{4.51E+02}	&	\textbf{1.00}	&	\textbf{5}	&	\textbf{Jul 2007}	&	\textbf{2, \textit{33}}	\\
GJ 494A	&	0.59	&	0.53	&	2.85	&	0.092	&	6.79E+13	&	9.99E+03	&	0.12	&	1.70E+04	&	0.91	&	8	&	2007	&	2, \textit{33}	\\
…	&	…	&	…	&	…	&	…	&	1.95E+14	&	1.09E+04	&	0.27	&	1.44E+04	&	0.89	&	…	&	2008	&	…	\\
GJ 388	&	0.42	&	0.38	&	2.24	&	0.047	&	9.77E+13	&	4.45E+04	&	0.97	&	3.97E+02	&	0.25	&	8	&	2007	&	2, \textit{34}	\\
…	&	…	&	…	&	…	&	…	&	1.11E+14	&	4.33E+04	&	0.92	&	1.78E+03	&	0.08	&	…	&	2008	&	…	\\
EQ Peg A	&	0.39	&	0.35	&	1.06	&	0.02	&	4.38E+14	&	1.81E+05	&	0.71	&	2.41E+04	&	0.29	&	4	&	Aug 2006	&	2, \textit{34}	\\
EQ Peg B	&	0.25	&	0.25	&	0.4	&	0.005	&	4.25E+14	&	2.14E+05	&	0.95	&	2.70E+03	&	0.42	&	8	&	Aug 2006	&	2, \textit{34}	\\
GJ 873	&	0.32	&	0.3	&	4.37	&	0.068	&	1.33E+15	&	3.45E+05	&	0.28	&	3.05E+04	&	0.61	&	8	&	2006	&	2, \textit{34}	\\
…	&	…	&	…	&	…	&	…	&	2.12E+14	&	2.82E+05	&	0.30	&	4.12E+03	&	0.20	&	…	&	2007	&	…	\\
GJ 9520 	&	0.55	&	0.49	&	3.4	&	0.097	&	1.76E+14	&	1.63E+04	&	0.85	&	4.44E+03	&	0.87	&	8	&	2007	&	2, \textit{33}	\\
…	&	…	&	…	&	…	&	…	&	2.38E+14	&	1.33E+04	&	0.63	&	7.56E+03	&	0.91	&	…	&	2008	&	…	\\
V374 Peg	&	0.28	&	0.28	&	0.45	&	0.006	&	1.29E+14	&	5.30E+05	&	0.82	&	1.43E+04	&	0.01	&	10	&	2005	&	34, 2, \textit{35}	\\
…	&	…	&	…	&	…	&	…	&	8.11E+13	&	3.86E+05	&	0.81	&	1.12E+04	&	0.01	&	…	&	2006	&	…	\\
GJ 1111	&	0.1	&	0.11	&	0.46	&	0.0059	&	1.84E+13	&	1.29E+04	&	0.79	&	6.54E+02	&	0.68	&	6	&	2007	&	2, \textit{36}	\\
…	&	…	&	…	&	…	&	…	&	9.22E+12	&	5.43E+03	&	0.31	&	1.26E+03	&	0.88	&	…	&	2008	&	…	\\
…	&	…	&	…	&	…	&	…	&	1.57E+13	&	4.47E+03	&	0.66	&	1.81E+03	&	0.79	&	…	&	2009	&	…	\\
GJ 1156	&	0.14	&	0.16	&	0.49	&	0.0081	&	1.39E+12	&	4.38E+03	&	0.02	&	3.74E+02	&	0.19	&	6	&	2007	&	2, \textit{36}	\\
…	&	…	&	…	&	…	&	…	&	2.55E+13	&	1.44E+04	&	0.12	&	1.69E+03	&	0.58	&	…	&	2008	&	…	\\
…	&	…	&	…	&	…	&	…	&	3.86E+11	&	1.04E+04	&	0.01	&	4.88E+02	&	0.05	&	…	&	2009	&	…	\\
\textbf{GJ 1245B}	&	\textbf{0.12}	&	\textbf{0.14}	&	\textbf{0.71}	&	\textbf{0.011}	&	\textbf{2.57E+13}	&	\textbf{3.34E+04}	&	\textbf{0.06}	&	\textbf{5.30E+03}	&	\textbf{0.37}	&	\textbf{4}	&	\textbf{2006}	&	\textbf{2, \textit{36}}	\\
…	&	…	&	…	&	…	&	…	&	4.89E+12	&	4.68E+03	&	0.13	&	4.86E+02	&	0.30	&	…	&	2008	&	…	\\
\textbf{WX UMa}	&	\textbf{0.1}	&	\textbf{0.12}	&	\textbf{0.78}	&	\textbf{0.01}	&	2.28E+15	&	2.19E+06	&	0.93	&	4.84E+04	&	0.38	&	\textbf{4}	&	2007	&	\textbf{2, \textit{36}}	\\
…	&	…	&	…	&	…	&	…	&	\textbf{3.89E+14}	&	\textbf{2.15E+06}	&	\textbf{0.85}	&	\textbf{3.33E+04}	&	\textbf{0.03}	&	…	&	\textbf{2008}	&	…	\\
\hline
\end{tabularx}
\begin{flushleft} 
1: \citet{Marsden2014}; 2: \citet{Vidotto2014}; 3: Petit et al. (in prep); 4: \citet{Saar1999}; 5: \citet{Hempelmann2016}; 6: \citet{DoNascimento2016}; 7: \citet{Jeffers2014}; 8: \citet{Petit2008}; 9: \citet{Morgenthaler2011}; 10: \citet{Fernandes1998}; 11: Jeffers et al. (in prep); 12: \citet{Valenti2005}; 13: \citet{Mosser2009}; 14: \citet{Boro2016}; 15: \citet{Boro2015}; 16: \citet{Maggio2000}; 17: \citet{Innis1988}; 18: \citet{Donati2003}; 19: \citet{Folsom2016}; 20: \citet{Takeda2007}; 21: \citet{Fares2009}; 22: \citet{Catala2007}; 23: \citet{Donati2008b}; 24: \citet{Fares2013}; 25: \citet{Udry2003}; 26: \citet{Melo2007}; 27: \citet{Ge2006}; 28: \citet{Udry2000}; 29: \citet{Simpson2010}; 30: \citet{Fares2012}; 31: \citet{Bouchy2005}; 32: \citet{Fares2010}; 33: \citet{Donati2008}; 34: \citet{Morin2008b}; 35: \citet{Morin2008}; 36: \citet{Morin2010} \\
\end{flushleft}
  \label{table:StellarSample}
\end{table*}
\end{center}

\section*{Acknowledgements}

The authors would like to thank the referee for thoughtful and constructive comments. MJ and KL acknowledge support from Science and Technology Facilities Council (STFC) consolidated grant number ST/R000824/1. VS acknowledges funding from the European Research Council (ERC) under the European Unions Horizon 2020 research and innovation programme (grant agreement no. 682393, AWESoMeStars). JFD acknowledges funding from the European Research Council (ERC) under the H2020 research and innovation programme (grant agreement 740651 NewWorlds). RF acknowledges funding from United Arab Emirates University (UAEU) startup grant number G00003269.  

\section*{Data Availability}

The data used to make Figures \ref{fig:slope}, \ref{fig:excess} and \ref{fig:confusagram} are in Table \ref{table:StellarSample} and can be accessed at  \url{https://doi.org/10.17630/4dea9f0c-9626-4655-ba61-8af5c05fbe38}. Archival data underpinning the plots is available at polarbase (\url{http://polarbase.irap.omp.eu}). 




\bibliographystyle{mnras}
\bibliography{refs} 







\bsp	
\label{lastpage}
\end{document}